# Network of Low-cost Air Quality Sensor for Monitoring Indoor, Outdoor, and Personal PM$_{2.5}$ Exposure in Seattle during the 2020 Wildfire Season

Jiayang He[1*], Ching-Hsuan Huang[2], Nanhsun Yuan[1], Elena Austin[2], Edmund Seto[2], Igor Novosselov [1,2]

[1] Department of Mechanical Engineering, College of Engineering, University of Washington, Seattle, WA, United States

[2] Department of Environmental and Occupational Health Sciences, School of Public Health, University of Washington, Seattle, WA, United States

* Corresponding author

E-mail address: jh846@uw.edu

# Abstract

The increased frequency of wildfires in the Western United States has raised public awareness of the impact of wildfire smoke on air quality and human health. Exposure to wildfire smoke has been linked to an increased risk of cancer and cardiorespiratory morbidity. Evidence-driven interventions can alleviate the adverse health impact of wildfire smoke. Public health guidance during wildfires is based on regional air quality data with limited spatiotemporal resolution. Recently, low-cost air quality sensors have been used in air quality studies, given their ability to capture high-resolution spatiotemporal data. We demonstrate the use of a network of low-cost particulate matter (PM) sensors to gather indoor, outdoor, and personal $PM_{2.5}$ exposure data from seven locations in the urban Seattle area, along with a personal exposure monitor worn by a resident living in one of these locations during the 2020 Washington wildfire event. The data were used to determine PM concentration indoor/outdoor (I/O) ratios, PM reduction, and personal exposure levels. The result shows that locations equipped with high-efficiency particulate air (HEPA) filters and HVAC filtration systems had significantly lower I/O ratios (median I/O = 0.43) than those without air filtration (median I/O = 0.82). The median $PM_{2.5}$ reduction for the locations with HEPA is 58 % compared to 20% for the locations without HEPA. The outdoor PM sensors showed a high correlation to the nearby regional air quality monitoring stations ($R^2$ = 0.93). The personal monitor showed high variance in PM measurements as the user moved through different microenvironments and could not be fully characterized by the network of indoor or outdoor monitors. The findings imply evidence-based interventions can be developed for reducing pollution exposure based on the combination of indoor, outdoor sensors. Personal exposure monitoring in individuals' breathing zones provided the highest fidelity data capturing temporal spikes in PM exposure.

# 1. Introduction

Climate-change-related wildfires have become more frequent and intense in the Western United States. Summer wildfire seasons are 40 to 80 days longer on average than they were 30 years ago [1]. Evidence suggests that California and other Western states will likely see ever-worsening fires for the coming century due to climate change and land management practices [2-5]. The intensified wildfires will release more smoke into the atmosphere [6], traveling significant distances [7]. Fine particulate matter ($PM_{2.5}$), a major pollutant found in smoke from wildfires, can travel deep into the respiratory tract [8]. The combustion-generated aerosols consist of elemental carbon and organic carbon fraction, which may be more toxic than other $PM_{2.5}$ sources and may have long-lasting impacts on health [9-12]. Complex flow structures associated with large-scale flames and low flame temperature in biomass burning lead to low carbonization of organic carbon, thus -- high levels of potentially carcinogenic polycyclic aromatic compounds [13-16]. Exposure to $PM_{2.5}$, particularly combustion-generated aerosols, has been linked to adverse respiratory and cardiovascular health effects, including ischemic heart disease, stroke, cardiovascular mortality, and exacerbations of asthma and chronic obstructive pulmonary disease [17-21]. More recently the wildfire PM exposures have been linked to higher severity and mortality of SARS-CoV-2 [22-25].

A series of large wildfires impacted air quality in western regions of the United States in 2020. The episode measured in this study (2020 Washington Labor Day fires) began on September 7, 2020, and were 90% contained by September 22. The fires burned over 41,000 acres of the forest [26]. Due to the SARS-CoV-2 pandemic shelter-in-place order by Washington state since early 2020, people probably spent a significant amount of time indoor during the 2020 wildfire season. The public health advice for protection from wildfire smoke exposure is to stay indoors, preferably in a "clean room" with filtered air, closed windows and doors, and minimize physical exertion. However, studies have shown that $PM_{2.5}$ could penetrate indoors even with all the windows and doors closed [27, 28]. With limited access to portable air cleaners during wildfires and increasing awareness of the health impact from wildfire smoke exposure, monitoring indoor $PM_{2.5}$ is critical to estimate household members' wildfire smoke exposure. Failure to assess the exposure to wildfire smoke could lead to misclassification of exposure in future epidemiology studies and have important public health implications for targeting smoke reduction interventions. The opportunity exists to improve personal exposure assessment and design individualized intervention strategies that would significantly reduce the adverse impact of PM pollution on human health, including the severity and mortality of Covid-19 cases [29, 30].

Recent advancements in low-cost particulate matter (PM) sensors led to their extensive use in various applications, such as air quality (AQ) monitoring in indoor [31-34] and outdoor [35-39] environments, including large-scale deployments [40-43]. Optical PM sensors rely on elastic light scattering providing size-resolved PM concentrations in the $0.3 - 10.0$ µm range. The low-cost sensor measurements may suffer from sensor-to-sensor variability due to a lack of quality control and differences between individual components.[44, 45] The scattering light intensity depends on particle size, morphology, complex index of refraction (CRI), and sensor geometry. [46] CRI sensitivity can be addressed by optimizing the design to measure scattered light at multiple angles simultaneously or by employing dual-wavelength techniques. [47, 48] However, these solutions are complex and involve expensive components that are not suitable for compact, low-cost devices. [31]

Environmental conditions were reported to affect sensor output, e.g., a non-linear response has been reported with increasing RH. [49-53] High humidity (RH > 75%) creates challenges for particle instruments; e.g., significant variations were observed between different commercially available instruments, such as Nova PM sensor [49] and personal DataRAM. [51] In addition, the RH measurement approach could also affect the sensor output [49, 50], e.g., the RH measurement based on reference monitoring site rather than inside the sensor enclosure may be different due to the microenvironment and transient effects. The selection of reference instruments with different measuring principles may also influence the calibration of low-cost sensors. For example, the calibration of the Plantower PM sensor in Jayaratne et al., 2018 was based on the tapered element oscillating microbalance (TEOM), while Zusman et al., 2020 calibrated the same sensor against the beta attenuation monitor (BAM) and federal reference method (FRM) measurements [50, 54]. The integrated mass measurements cannot account for temporal particle size and concentration variation during the calibration experiment. The instruments that directly measure aerosol size and concentration, such as aerodynamic particle sizer (APS), can be a better fit for sensor calibration.[44, 55]

As low-cost sensors find applications in pollution monitoring, various studies have evaluated the performance of low-cost PM sensors in laboratory and field settings.[44, 54, 56-61] These reports show that low-cost sensors yield usable data when calibrated against research-grade reference instruments.[42, 62, 63] The low-cost sensor networks have the potential to provide high spatial and temporal resolution, identifying pollution sources and hotspots, which in turn can lead to the development of intervention strategies for exposure assessment and intervention strategies for susceptible individuals. Time-resolved exposure data from wearable monitors can be used to assess individual exposure in near real-time [64].

This study utilized a network of indoor, outdoor, and wearable low-cost air quality sensors to evaluate 1) the effectiveness of intervention strategies used in different households in terms of $PM_{2.5}$ I/O ratios and $PM_{2.5}$ reduction during the 2020 Washington wildfire in seven locations, including residential and office buildings; 2) estimation of personal exposure by (i) wearable sensor and (ii) a combination of indoor and outdoor monitors, where the fraction of personal exposure from different microenvironments is determined based on the Global Positioning System (GPS) and time-resolved PM sensor data.

## 2. Methods

### 2.1 PM Monitor

The monitor used in this study consists of a PM sensor, a temperature/humidity/pressure sensor, a GPS module (ublox SAM-M8Q), and a display. The PM sensor (Plantower PMS A003, Beijing Plantower Co., Ltd, China; referred to as PMS hereafter) is an optical scattering-based sensor with a photodiode positioned normal to the excitation beam. The scattering light intensity is converted to a voltage signal to estimate PM number concentration and mass concentration using a proprietary calibration algorithm. The PMS provides estimated particle counts in six size bins with the optical diameter in 0.3-10 μm (#/0.1L) range and mass concentration (μg/m$^3$) for PM$_1$, PM$_{2.5}$, and PM$_{10}$. The mass concentrations can be set to "standard" and "atmospheric", altering the assumed particle density. The "standard" condition is designed to be used in industrial settings, whereas the "atmospheric" condition best measures particles in the ambient environment. When collecting the data, the "ATM" setting for PM$_{2.5}$ concentration was used; the sampling interval was set to be 10 seconds. In the analysis, our own calibration was developed. This was motivated by a recent report showing that including PM-specific terms (such as PM CRI and density) in the calibration algorithm improves the correlation between the data from Plantower PMS A003 sensor with reference particle counter for both number density and mass concentration. In contrast, the RH term did not improve calibration in the range of RH=17-80%. [62] Thus, the specific calibration for this wildfire event was developed and used in the analysis. The same device was also used as the wearable persona monitor. GPS data were used to coordinate the personal data to a specific location and attribute the PM exposures to the user's microenvironment.

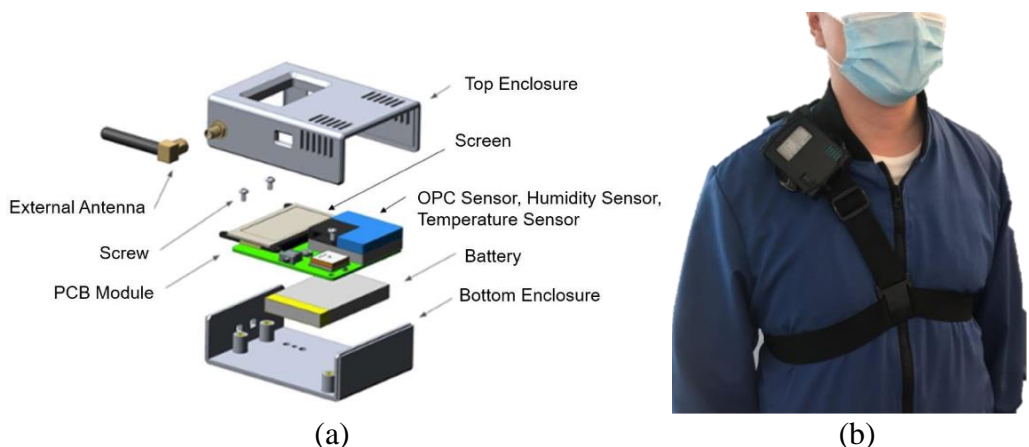

(a)                          (b)

**Figure 1.** a) Exploded view of the monitor enclosure; b) The wearable monitor.

## 2.2 Sampling Sites

The monitors were deployed in seven urban Seattle locations (see Figure A1). Each location had one outdoor sensor and at least one indoor sensor. The L2 location had two indoor sensors. One user from the sampling site L2 wore an additional personal monitor for the duration of the study. The study covered the wildfire episode between September 10 and September 21, 2020. The sampling sites included two University of Washington (UW) buildings and five residences in Seattle. Data from the nearby Puget Sound Clean Air Agency (PSCAA) regional stations were used for the sensor calibration. Before the study, information about the sites such as housing type, size, HVAC, primary indoor PM sources, and locations of the sensors were collected (see Table 1). Three sites (L1-L3) had portable air purifiers or built-in high-efficiency air filtration in HVAC systems. Location 2 (L2) had two indoor monitors in separate rooms. Both UW buildings (L6 and L7) were largely unoccupied due to the shelter-in-place order and were not equipped with HEPA filtration units at the time. The residents at the sampling sites were not given specific instructions on whether to keep the windows and doors open or closed but were asked about this after the sampling was completed. The research staff performed the sampling at the UW buildings, and the windows were kept closed during sampling.

**Table 1.** General characteristics of the sampling sites.

| Location ID | Building Type | Size (sq.ft) | HVAC | HEPA | Window Opening [a] | Indoor PM Sources [a] |
|---|---|---|---|---|---|---|
| L1 | 1-story SFH | 1600 | N | Y | Sometimes | Occasional cooking |
| L2-a L2-b | 1-story SFH | 1500 | N | Y | No | Occasional cooking |
| L3 | 2-story SFH | 3500 | Y[b] | N | No | Occasional cooking |
| L4 | 2-story SFH | 3000 | N | N | Always | Frequent cooking |
| L5 | Apartment | 800 | N | N | Sometimes | Occasional cooking |
| L6 | Office | 135 | Y | N | No | N/A |
| L7 | Office | 144 | Y | N | No | N/A |

[a] Self-reported information
[b] Electrostatic precipitator built in the HVAC
Definition of abbreviation: SFH = single-family home; Y = Yes; N= No; sq.ft = square feet

## 2.3 Data Analysis

The low-cost sensor data were corrected against the average of $PM_{2.5}$ from the two nearby PSCAA regional monitors. The correction model was generated using a data subset from the outdoor sensor outside a UW building (L7) during the wildfire event. L7 was largely unoccupied during the wildfire with minimum local activities that may influence $PM_{2.5}$ measurement compared to the other sampling sites. L7 is located at least 200 meters from any major traffic arterials. Sensor to sensor difference was within 10%, as shown in our previous aerosol chamber experiments [65]. Informed by our previous PMS sensor calibration study, a linear model or a quadratic were tested to fit using the outdoor sensor data from L7:

$$Ref = \beta_0 + \beta_1 \cdot PMS \qquad (1)$$

$$Ref = \beta_0 + \beta_1 \cdot PMS + \beta_2 \cdot PMS^2 \qquad (2)$$

where $\beta_0$, $\beta_1$, and $\beta_2$ are the regression coefficients, $Ref$ is the hourly reference $PM_{2.5}$ concentrations from the nearby PSCAA monitoring stations, and $PMS$ and $PMS^2$ are the linear and quadratic coefficients of the raw $PM_{2.5}$ data from the sensor, respectively. The fits with zero intercept ($\beta_0 = 0$) and non-zero intercept ($\beta_0 \neq 0$) were tested. The Bayesian Information Criterion (BIC) was used to select the optimal calibration model.

The time-resolved PM concentration I/O ratio was calculated to assess the smoke infiltration. We conducted the Wilcoxon signed-rank tests (for paired comparison) to compare the I/O ratio during the wildfire to the I/O ratio post the wildfire. To assess the reduction in PM levels, we compared indoor and outdoor time-resolved PM concentrations. We calculated $PM_{2.5}$ reduction for each site and the personal exposure as:

$$\textbf{PM}_{\textbf{2.5}} \textbf{ Reduction} = \frac{O-I}{O} \% \tag{3}$$

where $O$ is the average outdoor $PM_{2.5}$ concentration ($\mu g/m^3$) during the wildfire and $I$ is the average indoor or personal $PM_{2.5}$ concentration ($\mu g/m^3$).

To understand the contribution of each microenvironment to personal exposure, we attributed the user's daily $PM_{2.5}$ exposures in each location using GPS data with a 2.5m horizontal accuracy from the wearable monitor. The personal exposure attribution was done using Python 3.7.1. The raw $PM_{2.5}$ data were first aggregated into 10-min averages to reduce the data size without losing significant spatial resolution. The geocoordinates, recorded in conjunction with the $PM_{2.5}$ concentration, were grouped into three categories where the user spent most of the time: home, office, and other. We defined the buffer zones encircling the residence and the office locations with a 10-meter clearance to minimize misclassification error caused by the GPS drift. The home and office geolocation data were visually confirmed on a map for each occurrence when the user with the wearable monitor was at the location. Records collected outside these two buffer zones were classified as "other locations". Then the $PM_{2.5}$ exposures attributed to each microenvironment for each day was calculated as:

$$AC_{k_j} = \frac{C_{k_j} \times F_{k_j}}{\sum_{k=1}^{n} F_{k_j}} \qquad (4)$$

where $AC_{k_j}$ represents the attributable exposures of microenvironment $k$ to the total personal exposure on day $j$; $C_{k_j}$ is the hourly average PM$_{2.5}$ concentrations (μg/m$^3$) of microenvironment $k$ on day $j$ ; and $F_{k_j}$ is the fraction of time spent in microenvironment $k$ on day $j$.

## 3. Results and Discussion

### 3.1 PMS Sensor Correction

The data from L7 outdoor sensor shows a good agreement with the regional monitors with the pre-calibration R$^2$ = 0.92. The linear model shows the lower root-mean-square-error (RMSE) and BIC, with the overall RMSE improved from 18.47 μg/m$^3$ to 14.35 μg/m$^3$ against the regional monitors with the post-calibration R$^2$ = 0.94 (see Figure A2). The quadratic model did not result in a significant improvement; thus, the data from all the other sensors were corrected with the linear model. Note that while the PM$_{2.5}$ correction agreed with reference monitors using the linear model, the particle number density data from PMS were not evaluated for wildfire smoke. The particle size-dependent data from low-cost optical sensors are significantly influenced by the particle index of refraction, particle morphology, particle loading, sensor geometry, and environmental conditions [44, 46, 57, 66-68].

### 3.2 Time-resolved PM Concentrations

Figure 2 shows 1-hour averages of PM$_{2.5}$ concentrations measured by the sensor network and the regional monitors during the wildfire. The data are divided into indoor and outdoor categories. The indoor data are further divided into HEPA (L1-L3) and Non-HEPA (L4-L7) subsets. The shaded areas represent one standard deviation (1σ) of the measurements. The data from the outdoor monitors closely match the reference monitors. The correlation between the average outdoor PM concentration from the sensor network and the average of the reference monitors is 0.97.

The locations with active air filtration had a significantly lower PM concentration, while locations without HEPA filters had only slightly lower PM levels. Occasional spikes in PM concentration, above the already high baseline, were observed due to cooking activities. Note that even with the active PM control strategies implemented in several households, the average indoor $PM_{2.5}$ is still much higher than the typical (non-wildfire season) outdoor $PM_{2.5}$ levels ($< 10$ µg/m$^3$) in this region [69, 70].

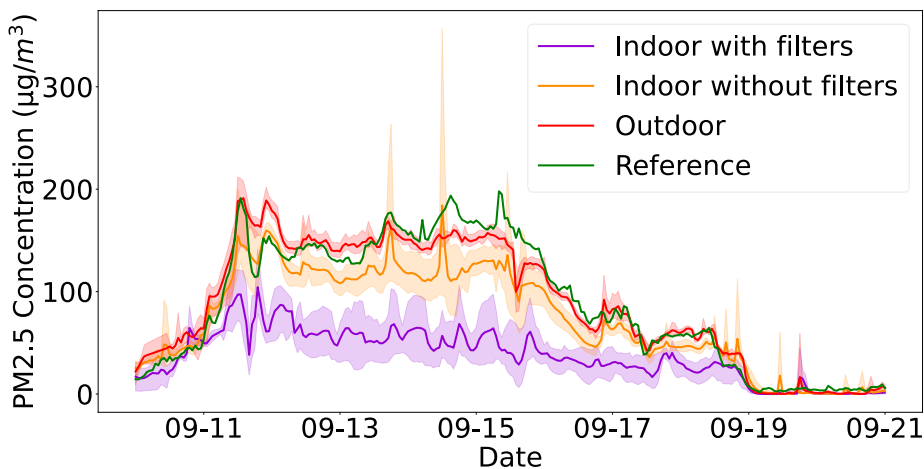

**Figure 2.** Corrected average indoor and outdoor $PM_{2.5}$ concentrations across the seven sampling sites compared to the reference monitors during the wildfire. The purple and orange lines illustrate the average indoor $PM_{2.5}$ concentrations at the sampling sites with and without HEPA filtration, respectively. The shading around each line shows the one standard deviation (1σ) of the measurement.

## 3.3 Detailed PM Concentrations and Analysis

We present a case-by-case analysis to provide insight into the effectiveness of aerosol mitigation strategies and the significance of the quantity and placement of sensors within the residences. Figure 3 shows the 1-hour average indoor, and outdoor data for all sites with the averaged $PM_{2.5}$ from the two reference monitors in green color for comparison.

Overall, the outdoor sensors are in close agreement with PSCAA data. In some cases, the outdoor sensors reported spikes in PM concentration likely due to the local activities near monitoring sites, which the PSCAA monitors could not detect (see Figure 3). These differences can be explained by (i) spikes in the local PM concentration or (ii) mismatch in sensor sampling rate as the network sensors sampling interval was set to ~ 10 seconds, while the reference monitors' reporting interval was 1 hour. Though both scenarios are possible, the PM levels difference between the locations suggests that some differences in outdoor $PM_{2.5}$ concentration were driven by local events that the regional monitors could not capture.

Only a moderate reduction in PM levels (16-29%) was observed when low-grade PM filers were used at L5, L6, and L7. Table 2 lists the mean and maximum of $PM_{2.5}$ for each sampling site. $PM_{2.5}$ reduction levels were calculated for each sampling location to indicate the effectiveness of mitigation strategies for households.

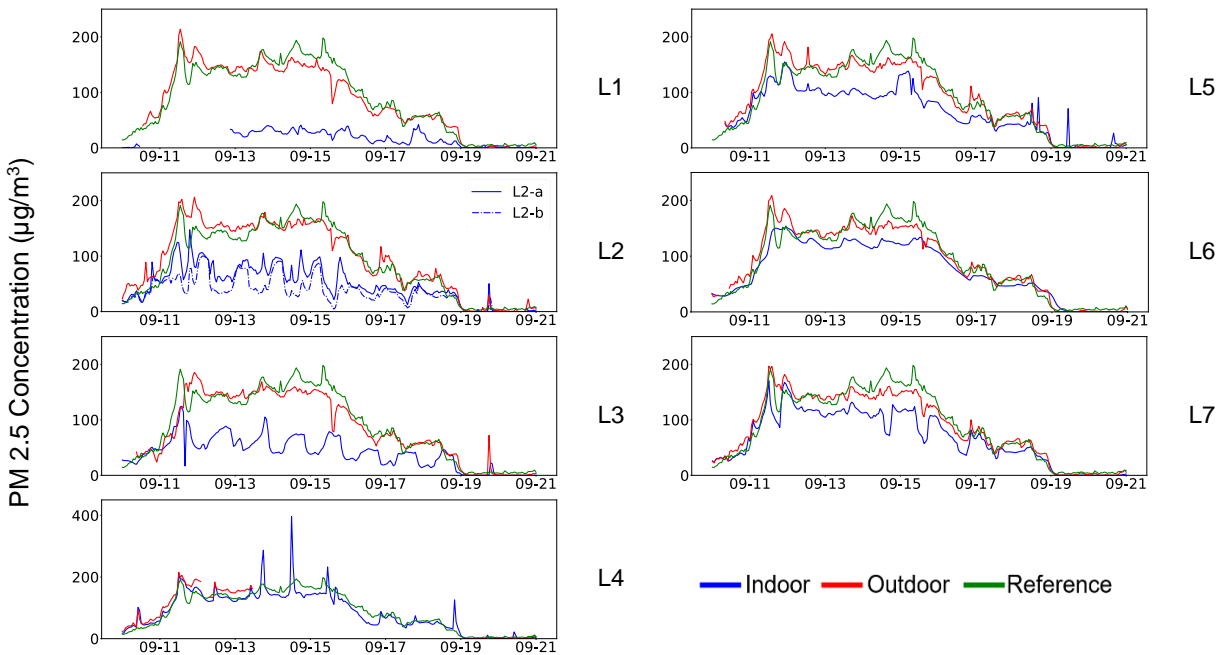

**Figure 3.** Time-series plots of indoor and outdoor $PM_{2.5}$ concentrations compared to the reference monitors for each sampling site during the wildfire. The blue and red lines represent the indoor and outdoor $PM_{2.5}$ measured by the sensors, and the green line represents the averaged $PM_{2.5}$ concentrations from the two nearby regional monitoring sites.

Location L1 had the indoor monitor placed in a relatively small room (home-office ~150 ft$^2$) with a high-volume HEPA filter for the entire wildfire episode. This relatively small "clean room" environment strategy resulted in the study's lowest median I/O ~ 0.2 (see Table A1). However, data for other locations (e.g., bedrooms, living room) within the residence are not available, which is problematic for assessing personal exposure. The I/O comparison is shown in Figure 4.

The residents of the L2 residence had the two monitors in separate rooms: L2a was placed in a larger living room (~350 ft$^2$) and L2b - in the adjacent home office (120 ft$^2$). Two portable HEPA filtration units were used: one in the living room, and the other was moved from the office to the bedroom at nighttime. The data from the living room has relatively low variance; however, the larger room was not cleaned as effectively as the smaller home office or bedroom, I/O ratio stayed relatively constant ~0.5. When the filtration unit was positioned in the office, the I/O ratios dropped to ~ 0.4. When the filter was moved from the home office to the bedroom, the PM level in the office increased to the level of the adjacent living room. The bedroom was not monitored by a fixed sensor; however, the resident's wearable sensor recorded a significant PM reduction in the bedroom during the night with the I/O ratio close to 0.1 (see Figure 5).

The L3 site had an electrostatic precipitator installed in the HVAC system. The HVAC system was controlled by a thermostat, which explains the periodic pattern of the PM concentration. The PM concentration (measured in the bedroom) dipped during the daytime when the forced air HVAC system was ON and went up during the nighttime when the HVAC was OFF.

L4 indoor monitor was placed in the kitchen where it detected the spikes from cooking activities in addition to the high background level from wildfire smoke intrusion. L4 residents kept their kitchen windows open during the wildfires, which explains the highest background indoor

PM level among all the sampling sites. L4 outdoor sensor stopped working three days after the deployment because it was accidentally unplugged from the AC outlet. I/O ratio for L4 was calculated using the data collected before September 13. The sampling site L4 without air filtration and closed windows had only a 13% indoor-outdoor difference.

The L5 residence did not have a central HVAC system. A portable AC unit (AeonAir Model #RPAC08EE) with a low-grade PM filter was used by the residents during the wildfire. The window in residence was closed for the entire duration of sampling (self-reported). L5 monitor was placed in the apartment's kitchen/living room area (200 ft2). PM concentration was lower than the outdoor level during the wildfire but higher than other residences with air filtration units. L6 and L7 are two UW buildings with HVAC systems but low-grade filtration units. The data is similar to the L5 residential site. The buildings were largely unoccupied during the wildfire due to COVID lockdown, and the windows were closed. The indoor $PM_{2.5}$ at L6 and L7 are lower than outdoor. The I/O ratios were similar to L5 (~0.7-0.8). Only a moderate difference in PM levels (16-29%) was observed when low-grade PM filers were used at L5, L6, and L7.

Interestingly, sites L3 and L7 with central filtration units show similar trends. PM concentration dipped rapidly when the HVAC was turned ON (controlled by a thermostat). As the central HVAC unit was OFF, the PM concentration climbed up due to infiltration of smoke. Though the analysis of HVAC performance is beyond the scope of this manuscript, these data can be used to design and optimize HVAC performance, such as the use of economizers, sensor-based controls, filter upgrades, etc.

**Table 2.** Summary of the indoor and outdoor $PM_{2.5}$ levels ($\mu g/m^3$) and I/O ratios for each sampling site.

| Location ID | Indoor | | Outdoor | | I/O ratios | | | | $PM_{2.5}$ Reduction (%)[a] |
|---|---|---|---|---|---|---|---|---|---|
| | Mean | Max | Mean | Max | Min | Median | Mean | Max | |
| L1 | 20.9 | 42.2 | 102.0 | 174.2 | 0.1 | 0.21 | 0.23 | 0.74 | 79.6% |
| L2-a | 58.7 | 147.4 | 114.5 | 206.3 | 0.18 | 0.52 | 0.54 | 1.41 | 48.7% |
| L2-b | 42.4 | 100.1 | 114.5 | 206.3 | 0.04 | 0.40 | 0.41 | 1.09 | 63.0% |
| L3 | 48.9 | 124.7 | 104.5 | 185.3 | 0.10 | 0.46 | 0.54 | 5.23 | 53.2% |
| L4 | 104.3 | 396.2 | 123.8 | 215.7 | 0.73 | 0.86 | 0.88 | 1.44 | 15.7% |
| L5 | 79.7 | 154.9 | 112.2 | 205.5 | 0.51 | 0.69 | 0.73 | 1.93 | 29.0% |
| L6 | 90.9 | 150.9 | 110.1 | 208.6 | 0.57 | 0.84 | 0.86 | 3.62 | 17.5% |
| L7 | 82.5 | 170.7 | 105.5 | 196.7 | 0.44 | 0.80 | 0.80 | 1.71 | 21.8% |

[a] Comparison between indoor and outdoor $PM_{2.5}$ levels during the wildfire calculated with equation 3.

Figure 4 compares the hourly $PM_{2.5}$ indoor/outdoor (I/O) ratios of different monitoring sites during the wildfire. The average I/O ratio across all seven sites was 0.62. The sites with HEPA filters (L1, L2-a, L2-b, and L3) and the sites without HEPA filters (L4- L7) had an average I/O ratio of 0.43 and 0.82, respectively, which are higher than the average I/O ratios found from another study conducted in the same region during the same wildfire episode (I/O=0.19 for the households with air cleaners; I/O=0.56 for the households without air cleaners) [71].

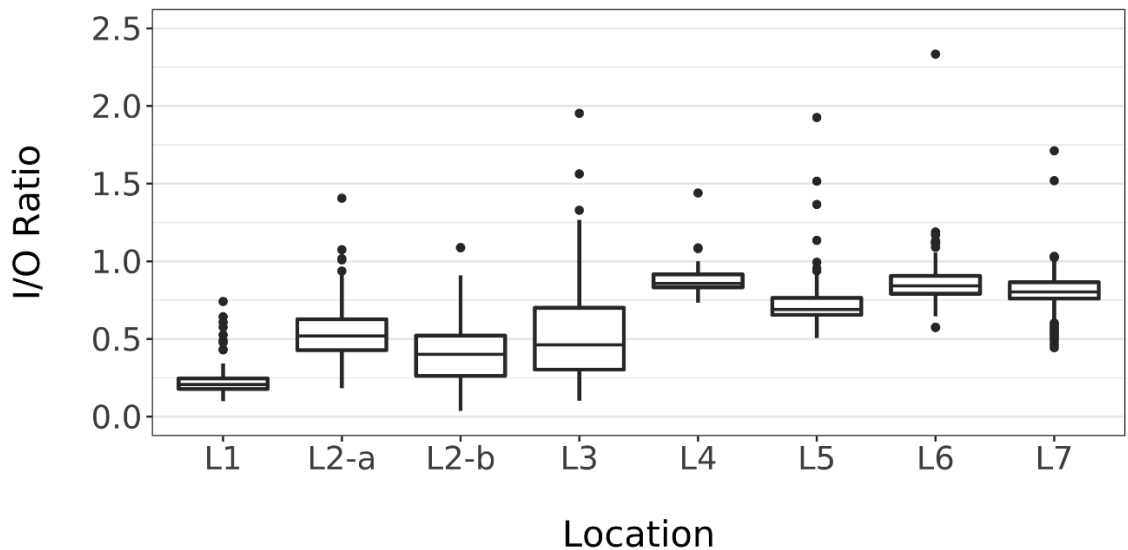

**Figure 4.** Boxplot of hourly I/O ratios for each sampling site.

## 3.4 Personal Exposure Measurement - A Case Study

To assess personal exposure as a function of the microenvironment, we compared the personal data measured by the wearable monitor with the wearer's home-based monitor data. Figure 5 shows the 10-minute average of $PM_{2.5}$ concentration measured by the user's personal and home-based monitors for reference. The wearable sensor PM$_{2.5}$ data showed a 68% PM$_{2.5}$ reduction, which is lower than the PM$_{2.5}$ reduction estimated using the wearer's indoor monitors. The shaded areas mark the nighttime (10:00 pm - 6:00 am). The personal monitor recorded significantly lower PM$_{2.5}$ concentration at night compared to the other two home-based monitors located in the living room and the home office, which can be explained by the colocation of the HEPA air cleaner and wearable monitor. The user is an aerosol researcher who monitored his exposure during the wildfire. The difference in the personal exposures, measured by the wearable and the home-based monitor, indicates that access to special hyperlocal resolution (Indoor - room level, Outdoor – 2.5 meters, Wearable sensor level – 2.5m with 10 m buffer) for PM concentration data  can enhance the efficacy of PM exposure interventions.

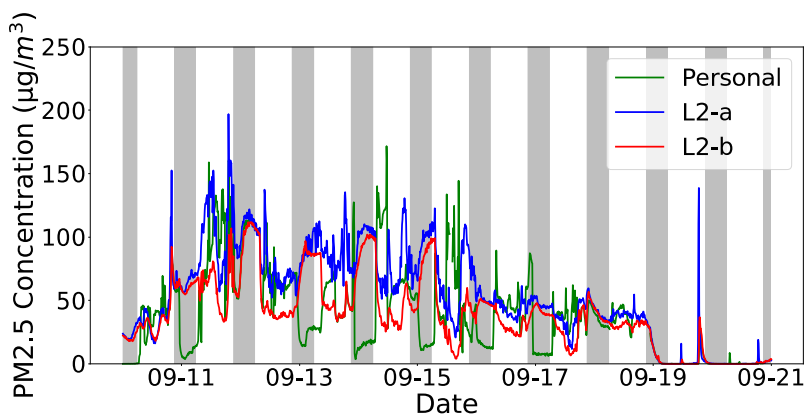

**Figure 5.** Profile of 10-minute averaged PM$_{2.5}$ concentrations by the personal monitor (green line) and the bedroom monitors (the blue and purple line) during the wildfire.

We also apportioned the user's exposure based on the GPS data. Figure 6a shows the 10-minute average of PM$_{2.5}$ concentration measured by the personal sensor color-coded based on the microenvironment. The user spent most of his time at home and with 15% time in the office and 9% in other locations. Exposure outside the home and the office was categorized into other locations. The wearable monitor recorded higher PM$_{2.5}$ levels in the user's workplace and other locations than at the residence. The work location did not have an efficient filtration system based on the geolocation and PM concentration data. Figure 6b shows the weighted daily average personal exposure in different microenvironments. The personal exposure contribution from the office and other locations was 36% of the total smoke exposure during the wildfire, while the time spent in these environments was 24% (see Figure A3). The attribution of personal exposure to microenvironments can help personal activities during wildfire seasons.

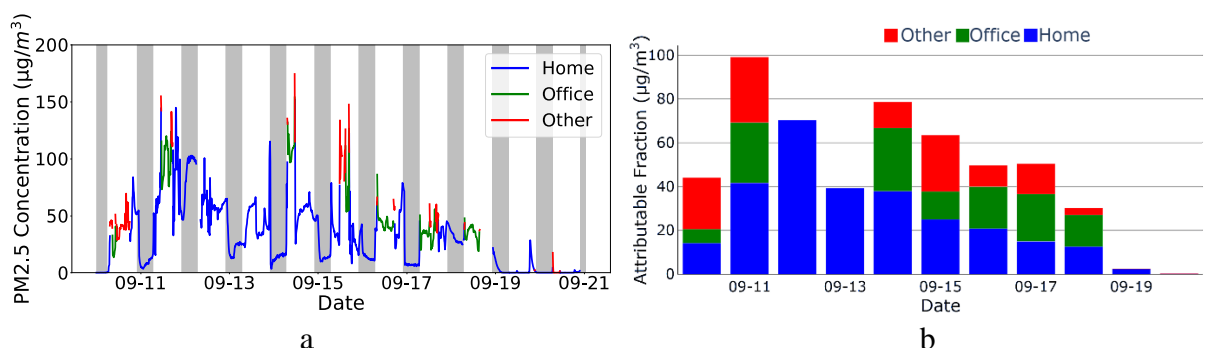

**Figure 6.** a) Time series plots of the personal PM$_{2.5}$ exposure color-coded based on the microenvironment the user was in; b) Weighted daily average personal exposure in different microenvironments.

## 4. Conclusions

This study demonstrates the application of a low-cost sensor network for air quality monitoring during the 2020 Washington wildfire event. The outdoor PM$_{2.5}$ data from the sensor network had an excellent agreement with the nearby PSCAA regional monitors. The spatial variance for PM$_{2.5}$ in the urban area was low during the wildfire event. Our results showed that during the 2020 Washington wildfire, the outdoor PM$_{2.5}$ level was as high as >200 μg/m$^3$ in the Seattle area. Using portable HEPA air cleaner was associated with lower indoor PM$_{2.5}$ levels during the wildfire episode, with the PM$_{2.5}$ reduction of 50 - 77 % among the sampling sites. However, the observed levels were still higher than the typical Seattle outdoor PM levels (< 10 μg/m$^3$). The I/O ratio was driven by the smoke infiltration and the quality of air filtration. The personal monitoring results highlighted the influence of microenvironments on an individual's exposure to $PM_{2.5}$. Although this study had a relatively small sample size, it demonstrated that personal action, such as staying indoors and using HEPA air cleaner, can reduce personal exposure to wildfire smoke. The personal exposure analysis suggests knowledge about PM levels can lead to a reduction in exposure. More extensive studies and a collection of time-activity information are warranted to investigate the source of $PM_{2.5}$ exposure and the health impact of PM exposure.